\documentclass[a4paper,11pt]{article}
\pdfoutput=1 
\usepackage[T1]{fontenc} 
\usepackage{amsmath, mathrsfs, amssymb,amsfonts,amsthm,graphicx, epsf, yfonts}

\def\Z#1{_{\lower2pt\hbox{$\scriptstyle#1$}}}
\def\vZ{v\Z{Z}} \def\vD{v\Z{D}} \def\vK{v\Z{K}}  \def\vS{v\Z{S}}
\def\vDr{v\Z{D,r}} \def\vDrr{v\Z{D,rr}} 
\def\vDSol{v\Z{D\odot}}
\def\dd{\mathop{\text{d}\!}}
\def\vKr{v\Z{K,r}}  

\title{\boldmath Effective galactic dark matter: first order general relativistic corrections}

\begin{document}

\maketitle

Federico Re$^{a,b}$, Marco Galoppo$^c$

$^a\;$ Dipartimento di Fisica \lq\lq Giuseppe Occhialini\rq\rq, Universit\`{a} di Milano Bicocca,\\ Piazza dell'Ateneo Nuovo 1, 20126, Milano, Italy \\
$^b\;$ INFN, sezione di Milano, Via Celoria 16, 20133, Milano, Italy \\
$^c\;$ School of Physical \& Chemical Sciences, University of Canterbury, \\ Private Bag 4800, Christchurch 8041, New Zealand

federico.re@unimib.it,
marco.galoppo@pg.canterbury.ac.nz

\medskip
\textbf{Abstract}. 
Stationary, axisymmetric, dust sourced solutions of Einstein’s equations
have been proposed as fully general relativistic models for disc galaxies. These models introduce a novel physical element, i.e., a non-negligible dragging vortex emerging from a full consideration of the essential self-interaction of matter and geometry in general relativity, which might demand a profound recalibration of the inferred amount of dark matter in disc galaxies. Within this framework, we identify the correct observables for redshift-inferred rotation curves of distant galaxies, correcting previously overlooked mistakes in the literature. We find that the presence of the dragging vortex introduces non-negligible corrective terms for the matter density required to maintain a stable physical system. We present the first estimate of the dragging speed which is required to explain a non-negligible fraction of dark matter in disc galaxies. In particular, we show that a sub-relativisitc dragging velocity of tens of kilometers per second in the neighbourhood of the Sun is sufficient to reduce the need of dark matter by 50\% in the Milky Way. Finally, we find that the presence of such a dragging vortex also returns a net contribution to the strong gravitational lensing generated by the galaxy. Thus, we show that the considered class of general relativistic galaxy models, is not only physcially viable, but suggests the need for recalibration of the estimated dark matter content in disc galaxies, with far reaching consequences for astrophysics and cosmology.

\section{Introduction}\label{sec:intro}

The most famous open problem of present-day astrophysics and cosmology is the \textit{Problem of the Missing Mass}. Namely, many astrophysical phenomena -- the rotation curves of disc galaxies \cite{RotSpiral1,RotSpiral2}; the virial estimates of galaxies in galaxy clusters; the gravitational lensing produced by such objects \cite{lensing}; the thermodynamic emission of X-rays in galaxy clusters \cite{gas}; the dynamical properties of the two Bullet Clusters \cite{bullet1,bullet2,bullet3} -- and other cosmological ones -- the multipolar spectrum of the cosmic microwave background \cite{CMB1,CMB2}; the formation of structures from initial matter inhomogeneities \cite{perturb1, perturb2}; and others -- require significantly more mass than the visible one to justify their dynamics. We will call all these types of observations and inferences as \lq\lq Missing Mass Phenomena\rq\rq~(MMP).

The most natural hypothesis to explain MMP is the presence of matter not detectable through electromagnetic interaction: the so-called dark matter (DM) \cite{Bertone:2018krk,profumo_introduction_2019}. About the exact nature of the DM, many speculations have been proposed: hypothetical particles with a big mass, interacting only via the weak force (the WIMPs, Weakly Interactive Massive Particles) \cite{WIMP1, WIMP2, WIMP3, WIMP4, WIMP5}; lighter particles whose field auto-interacts (the ALPs, Axion-Like Particles) \cite{ALP}, so that the dark matter halos would be the solitonic solutions of its field equation; or even macroscopic opaque objects (the MACHOs, MAssive Compact Halo Objects) \cite{MACHO1, MACHO2, MACHO3, MACHO4}. However, alternative explanations to MMP, avoiding the presence of actual matter, have been proposed.

Any such hypothesis that wishes to be an alternative to the DM paradigm can start from the general observation that \emph{all MMP have gravitational nature}.
This fact is obvious for the galaxy rotation curves or the Bullet Clusters. However, it also holds for mass measurements via gravitational lensing -- as the bending of space-time is a gravitational phenomenon -- and the inferred mean temperature of the gases in galaxy clusters -- essentially related to the estimation of the gas particles' velocities, and thus the Virial Theorem applied to a gravitational system. Analogous analyses can then be performed for any other MMP.

It can thus be pondered whether MMP point to a misunderstanding of how gravity works. Such claim constitutes the underlying hypothesis behind all the attempts to modify  Newtonian Gravity, leading to the so-called MOND (MOdified Newtonian Dynamics) \cite{Milgrom:1983ca, MOND1, MOND2} and MOG (MOdified Gravity) \cite{Bekenstein:2004ne, MOG1, MOG2} Theories. However, the currently accepted theory of gravity -- i.e., General Relativity (GR) -- already represents a stark departure from the Newtonian theory. Hence, we ask: \emph{can GR modifications be enough to justify the discrepancies between the expected and observed mass, up to a non-negligible fraction?}

The intuitive answer, for many astrophysicists, theoretical physicists, and cosmologists, is certainly: no. All the MMP are in a low-energy régime, exhibiting sub-relativistic dynamical speeds and weak gravitational forces, so that GR is intuitively conceived as almost identical to Newtonian gravity. In other words, a widespread belief about GR assumes that its low-energy limit coincides with the Newtonian theory. However, GR does allow for low-energy régime, totally non-Newtonian phenomena, whose understanding cannot be given by a mere correction on the Newtonian paradigm. An interesting example of non-Newtonian GR features is given by geons \cite{geons1,geons2}. These are non-trivial vacuum solutions, i.e. solitons, which represent localized gravitational waves, as they do not radiate in every direction but maintain their profile, held together by their gravitational energy. Such a space-time is in the low-energy régime, but it is clearly not Newtonian.

From a theoretical point of view, the non-Newtonian features of GR can be ascribed to deep differences between the two theories: the Newtonian equations are linear, whilst the Einstein Equations (EEs) are not; Einstein's gravitation field carries energy and momentum, in contrast to the non-dynamical Newtononian one; and finally GR has more degrees of freedom than the Newtonian theory (namely, the ten metric components, minus four gauges, compared to only one scalar gravitational potential). It is hence theoretically justified to investigate if the MMP involve dynamical systems for which the low-energy limit of GR does not coincide with the Newtonian description. This is what we call the paradigm of effective DM from GR. In particular, if we focus on disc galaxies, we could expect that the gravitational field would globally participate actively in the overall dynamics, by exchanging both energy and momentum with matter, according to the fully nonlinear nature of Einstein’s equations.

The first attempt to model disc galaxy dynamics in GR can be traced back to the work of Cooperstock and Tieu \cite{cooperstock_galactic_2007,cooperstock_general_2008}. However, these models were compounded by physical and mathematical pathologies, and therefore cannot be considered valid galaxy models in any cogent sense of the term \cite{vogt_presence_2005,Korzynski:2005xq}.  A later model, which sidestepped some of the unphysical features of the previous attempts, was proposed by Balasin and Grumiller \cite{BG}. Interestingly, the Balasin and Grumiller model (BG) has been proven to fit a rotation curve to the Milky Way, as inferred from the proper motion and redshifts of a sample of $7.2\times10^6$ young stars from the GAIA--DR3 survey, lying within $r<19\,$kpc of the galactic centre and $-1<z<1\,$kpc about the galactic plane, without any need for DM \cite{crosta2020,Crosta2023}. Nonetheless, BG still presents both curvature singularities and topological defects on the rotation axis \cite{BG,Costa1,Galoppo2}. Moreover, it has been proven to generate unphysical time delays between lensed images on the galactic plane \cite{Galoppo1}. Therefore, BG should at best be taken as a local model, describing the dynamics of a disc galaxy on a circular ring on the galactic plane. However, both BG and the Cooperstock--Tieu models depict rigidly rotating galaxies. As such, both models display a clearly unphysical feature -- galactic rigid rotation -- which, in our view, disqualifies them even as viable local models, regardless of the respective mathematical pathologies. 

To bypass the deficiencies introduced by unphysical rigid rotation, the physically viable class of differentially rotating models was introduced in \cite{Astesiano:2021ren}. The authors showed that introducing differential rotation further boosts the need for a recalibration of DM in disc galaxies, if these models can locally describe such rotationally supported astrophysical systems.
Further work on identifying the correct observables of such systems was then carried out in \cite{Re2023} and in \cite{AstesianoRuggiero1}.

Finally, we point out that a different line of research tries to neglect the nonlinearity of GR, developing models in the linearised régime; namely, the gravitomagnetic one \cite{Corre2015, Corre2017, Ludwig2021, Astesiano0, AstesianoRuggiero1, AstesianoRuggiero2, Srivastava2023, Corre2023a, Corre2023b, Corre2024}. These papers propose that a change in the boundary conditions at spatial infinity w.r.t. the conventional Minkowskian conditions, when studying the stationary approximation of a disc galaxy, would already be sufficient within the linear gravitomagnetic approximation to account for the observed rotation curves. The change in the boundary conditions would then be driven by the nonlinearities of GR effects in the formation of the galaxy, which would be accounted for precisely by the proposed effective change at spatial infinity. Following this interpretation, the field has produced interesting results which seem to well match the astrophysical observations (see e.g., \cite{Ludwig2021, AstesianoRuggiero1, AstesianoRuggiero2}).

In this paper, we focus on the application of the paradigm of effective DM from GR to the rotation curves of disc galaxies and their gravitational lensing. In Section \ref{sec:Metric}, we introduce the general spacetime metric, which describes the galactic plane geometry. Within Section \ref{sec:Observers} we discuss the correct physical observables for these systems and comment upon the region of validity of the differentially rotating solutions. In Section \ref{sec:Estimation} we give the first estimate of the strength of the dragging needed in a physical galaxy to give a non-negligible contribution in terms of effective DM. In Section \ref{sec:GravLensing} we show that the geometries considered give rise to DM-like effect in gravitational lensing. Section \ref{sec:conclusions} is dedicated to conclusions and future perspectives.

\section{General relativistic galaxy metric}\label{sec:Metric}

To model galaxies within the GR framework, let us consider stationary axisymmetric solutions of
the Einstein equations sourced by a pressureless fluid -- \emph{dust}. We consider an idealized version of the physical system in stationary motion and with exact symmetry around its rotation axis. We thus neglect the morphological evolution of the galaxy and any structure that breaks the axisymmetry (e.g., bars or spirals). Hence, spacetime possesses two Killing vectors, generating the time translation and the rotation around the axis, which allow a clear definition of the two coordinates $t$ and $\phi$. The parametrization of the spacetime is then completed by the axial and radial coordinates, respectively, $z$ and $r$. The metric takes the Lewis--Papapetrou--Weyl form
\begin{align}
    \label{eq:metric}
    ds^2 =-c^2 e^{2\Phi(r, z)/c^2}(dt-A(r, z)  \, d\phi)^2 +e^{-2\Phi(r, z)/c^2}\left[r^2  \, d\phi^2 +e^{2k(r, z)/c^2}(dr^2+dz^2)\right]\, .
\end{align}
where $\Phi(r,z)$ is related to the conventional Newtonian potential, $A(r,z)$ is related to frame-dragging, and $k(r,z)$ is a conformal factor on the 2--dimensional space of orbits of the isometry group generated by the Killing vectors.
The energy-momentum tensor takes the form
\begin{equation}
\label{eq:EMT}
    T_{\mu\nu}=c^2\rho(r, z) \, U_{\mu}U_{\nu}\, ,
\end{equation}
where $\rho(r,z)$ is the density of coarse-grained dust particles, each with a 4--velocity $U^\mu$, given by
\begin{equation}
\label{eq:U}
    U^{\mu}\partial_\mu=(-H)^{-1/2}\left(\partial_t +\Omega \, \partial_{\phi}\right)\, ,
\end{equation}
where $d\phi/dt=\Omega(r, z)$ uniquely defines the angular speed of rotation at any point, and $H(r, z)$ is a normalization factor. Since $U^{\mu}U_{\mu}=-c^2$, it follows that
\begin{equation}
\label{eq:Hnorm}
     H=-e^{2\Phi/c^2}(1+A \, \Omega)^2+e^{-2\Phi/c^2}r^2  \, \Omega^2/c^2\, .
\end{equation}
The general solution is thus characterized by six fields $(\Phi, A, k, \Omega, H, \rho)$, constrained by the EEs and the normalization condition \eqref{eq:Hnorm}. These, with the Minkowskian boundary condition at spatial infinity, set the mathematical problem. 

\subsection{The $(\eta,H)$ formalism}

The full coupling between matter and geometry has already been worked out \cite{stephani_kramer_maccallum_hoenselaers_herlt_2003}. The general solution to the EEs is found to be characterized by two fields, $\eta$ and $H$, depending only on the coordinates $(r,z)$. Using these two fields, the relevant quantities are written as 
\allowdisplaybreaks
\begin{align}
    & g_{tt}=c^2\frac{r^2\,\Omega^2/c^2-\left(H-\eta\,\Omega /c^2\right)^2}{-H} \label{eq:EHgtt} \, ,\\
    & g_{t\phi}=\frac{\eta^2/c^2-r^2}{-H}\, \Omega+\eta \label{eq:EHgtphi} \, ,\\
    & g_{\phi\phi}=\frac{r^2-\eta^2/c^2}{-H} \, ,  \label{eq:EHgphiphi} \\
    &\mu_{,r}=\frac{g_{tt,r}\,g_{\phi\phi,r}-g_{tt,z}\,g_{\phi\phi,z}  -g_{t\phi,r}^2+g_{t\phi,z}^2}{2c^2\,r} \,, \label{eq:EHmur}\\
    &\mu_{,z}=\frac{g_{tt,z}\,g_{\phi\phi,r}-g_{tt,r}\,g_{\phi\phi,z}-2\,g_{t\phi,z}\,g_{t\phi,r}}{2\,c^2r} \,, \label{eq:EHmuz} \\
    & 8\pi G e^{\mu/c^2} \rho = \frac{\eta_{,r}^2+\eta_{,z}^2}{4\eta^2 r^2}\left[\eta^2\left(2-\eta\ell(\eta)\right)^2 - c^4r^4\ell(\eta)^2\right] \, , \label{eq:eHdensity}
\end{align}
where it can be proven $H(r,z) = H(\eta)$, and we define $\ell = (dH/d\eta)/H$. Furthermore, we have $\Omega(r,z) = \Omega(\eta)$ from the differential relation
\begin{equation}
\label{H Omega}
     c^2\dd H=2\eta \dd \Omega\,. 
\end{equation}
Finally, the field $\eta(r,z)$ has to solve for the non-linear Partial Differential Equation (PDE)
\begin{align}
    &\left(\eta_{,rr}-\frac{1}{r}\eta_{,r}+\eta_{,zz}\right)\left(2-\eta\ell\right) + \left(\eta_{,r}^2-\eta_{,z}^2\right)\left(\eta\ell'-\ell\right)\left(1+\frac{r^2}{\eta^2}\right) + r^2\frac{\ell}{\eta}\left(\eta_{,rr}+\frac{3}{r}\eta_{,r}+\eta_{,zz}\right) = 0 \, .
    \label{eq:EtaEq}
\end{align}
Eq. \eqref{eq:EtaEq} uniquely determines $\eta(r,z)$ once $H(\eta)$ and $\eta(r,0)$ are arbitrarily assigned. Henceforth, we refer to this class of solutions -- whose every realization is entirely defined by the arbitrary choice of two functions of one variable -- as the $(\eta,H)$ class. 
Remarkably, by defining
\begin{align}
\label{new_eta_nonlinear}
    \Psi(r, z):=\eta +\frac{c^2}{2} \, r^2\int\frac{1}{\eta}\frac{dH}{H} -\frac{1}{2}\int\frac{\eta}{H} \, dH\, ,
\end{align}
we can cast the non-linear PDE \eqref{eq:EtaEq} into a linear, sourceless PDE, i.e., the Grad--Shafranov homogenous equation \cite{ruggiero_stationary_2024} 
\begin{equation}
    \Psi_{,rr}-\frac{1}{r}\Psi_{,r}+\Psi_{,zz}=0\, .
    \label{eq:GS}
\end{equation}

\subsection{Non-linear gravitomagnetism}
The class of metrics discussed to locally model galaxies in GR can be treated using a non-linear formalisation of gravitomagnetism, so that the EEs and the Equations of Motions for a test particle can be cast exactly, without the need for linear approximations. That is, following \cite{LandauLifshitz, NLGM1, Natario2007, costa_gravito-electromagnetic_2014, NLGM3, Costa1, Costa2}, we can define the gravitoelectric field $\mathbf{G}:=-\mathbf{\nabla}\Phi$ and the gravitomagnetic field $\mathbf{H}:=e^{\Phi/c^2}\left(\mathbf{\nabla}\times\mathbf{A}\right)$, in analogy with the electric and magnetic field, respectively. The EEs for a metric of the type \eqref{eq:metric} then become
\begin{equation}
\label{eq:nlGM}
    \begin{cases}
        \mathbf{\nabla}\cdot\mathbf{G}=\mathbf{G}^2/c^2+\mathbf{H}^2/2c^2-4\pi G\rho \\
        \mathbf{\nabla}\times\mathbf{G}=0 \\
        \mathbf{\nabla}\cdot\mathbf{H}=-\mathbf{G}\cdot\mathbf{H}/c^2 \\
        \mathbf{\nabla}\times\mathbf{H}=2\mathbf{G}\times\mathbf{H}/c^2-16\pi G\mathbf{j}/c^3
    \end{cases}\, ,
\end{equation}
where we have defined $j^{\mu}:=-T^{\mu\nu}U_{\nu}/c^2$ and on the equatorial plane we have $\mathbf{G}(\mathbf{x})=-G(r, z)\hat{r}$ and $\mathbf{A}(\mathbf{x})\approx-[\vD(r, z)/c^2]\hat{\phi} \Rightarrow \mathbf{H}(\mathbf{x})=-H(r, z)\hat{z}$. We note that, even within the non-linear gravitomagnetism formalism, the geodesic equation for a massive test-particle with four-velocity $\dd x^\mu/\dd \tau = u^\mu$, where $\tau$ is its proper time, can be understood in terms of a general relativistic analogue of the Lorentz force (see e.g., \cite{Costa1,Costa2}), i.e.,
\begin{equation}
    \frac{\dd \mathbf{u}}{\dd \tau} = \gamma\left[\gamma \mathbf{G} + \mathbf{u} \times \mathbf{H} \right] = \frac{1}{m}\mathbf{F_{\text{GEM}}}\, ,
\end{equation} 
where $\gamma:= U^\mu u_\mu$, with $u_\mu = e^{-\Phi/c^2}\partial_t$. Here, we have to stress that the non-linearities in \eqref{eq:nlGM} allow for solutions that are not just second-order perturbations of the solutions of the usually employed linear gravitomagnetism. For instance, $\textbf{H}$ is not forced to be directed towards the positive $\hat{z}$ direction, although $\textbf{j}$ is everywhere parallel to $\hat{\phi}$. We will see that whilst the $(\eta,H)$ formalism is better suited to estimate the amount of effective DM introduced by the GR paradigm, the non-linear gravitomagnetism approach is superior in dealing with lensing calculations.

\subsection{Physical domain of validity of the models}\label{sec:PhysicalValidity}

In our opinion, the $(\eta,H)$ class can be used to model the average disc galaxy dynamics. However, these models still present some drawbacks, which limit their domain of applicability. 
Indeed, the solutions under discussion are \emph{dust} models, devoid of pressure support. As such, we do not expect them to appropriately describe the physical system far away from the galactic plane along the $z$ direction.

This can be better understood by comparing the $(\eta, H)$ models with their Newtonian analogous. If a pressureless fluid in stationary rotation is substituted in the laws of Newton's dynamics, the balance along the $z$ axis returns $v_{,z}=0,\, \rho_{,z}=0$. In other words, the symmetries in $t$ and $\phi$ force a cylindrical symmetry, so that the only stable system is an infinite cylinder, otherwise the matter would collapse on the asymmetries along $z$. The other stable, although singular, solution is the thin-razor disc; namely, the case in which the matter has already collapsed on the $z=0$ plane. The physical system of a disc galaxy with a small, although not zero, thickness needs some amount of effective pressure, since the thickness can only be sustained by such pressure.

The $(\eta,H)$ models do not include any pressure in their description of the physical system. Therefore, they must require a rough coarse-graining scale along the $z$ axis to be applied to the study of disc galaxies. These solutions are to be taken as effective models, reliable in the study of the average thin disc dynamics for disc galaxies. They should not be applied far away from the galactic plane, where they would return unreliable physical behaviour for the real system. 

\section{Relevant observers}\label{sec:Observers}
Physical measurements are defined by relevant classes of observers. We thus begin by first introducing the Zero Angular Momentum Observer (ZAMO) coframe \cite{ZAMO1,ZAMO2}
\begin{align}
{\mathbf \omega}\Z Z=\Bigl(\frac{r\dd t}{\sqrt{g_{\phi\phi}}}\,,\sqrt{g_{\phi\phi}}\left(\dd \phi \,-\,\chi \dd t\right)\,, e^{(k-\Phi)/c^2} \dd r\,,e^{(k-\Phi)/c^2} \dd z\Bigr)\,, \label{eq:Zcoframe}
\end{align}
where $g_{\phi\phi}=r^2 e^{-2\Phi/c^2}-c^2 A^2 e^{2\Phi/c^2}$, with $\chi= - g_{t\phi}/g_{\phi\phi}$ and $g_{t\phi}=- c^2\,A\,e^{2\Phi/c^2}$. The dual ZAMO tetrad frame is then
\begin{align}
{\mathbf e}\Z Z=\Bigl(\frac{\sqrt{g_{\phi\phi}}}{r}&\left(\partial_t\,+\,\chi\partial_\phi\right)\,,\frac{1}{\sqrt{g_{\phi\phi}}}\partial_\phi\,, e^{-(k-\Phi)/c^2} \partial_r\,,e^{-(k-\Phi)/c^2} \partial_z\Bigr)\,.\label{eq:Zframe}
\end{align}
Starting from the formula for the dust particles' velocity measured by ZAMOs,  $\vZ =({{{\mathbf e}\Z Z}^1}\cdot{\mathbf U})/({{{\mathbf e}\Z Z}^0}\cdot{\mathbf U})$, we find the physically relevant relation
\begin{equation}
    \label{eq:eta}
    \eta(r,z) = r\, \vZ (r,z) \, .
\end{equation}
Eq. \eqref{eq:eta} gives us the physical interpretation of the field $\eta$, this is the angular momentum density per unit of mass measured by ZAMO. Now, let us define the effective Lorentz factor $\gamma\Z Z:= {{\mathbf e}\Z Z}^0\cdot{\mathbf U}$, so that 
\begin{equation}
    \label{eq:vZRelation}
     -H \,\gamma_\mathrm{Z}^2 \,\vZ=r\,(\Omega-\chi)\,.
\end{equation}
The second congruence of relevance are ideal Stationary Observers (SOs), with coframe \cite{Costa1}
\begin{align}
{\mathbf \omega}\Z S=\Bigl(e^{\Phi/c^2}&\left(\dd t \,+\,A \dd \phi\right)\,, r\,e^{\Phi/c^2} \dd \phi\,, e^{(k-\Phi)/c^2} \dd r\,,e^{(k-\Phi)/c^2} \dd z\Bigr)\,, \label{eq:Scoframe}
\end{align}
The dual SO tetrad frame is then
\begin{align}
{\mathbf e}\Z S=\Bigl(e^{-\Phi/c^2}&\partial_t\,,\frac{e^{\Phi/c^2}}{r}\left(\partial_\phi-A\,\partial_t\right)\,, e^{-(k-\Phi)/c^2} \partial_r\,,e^{-(k-\Phi)/c^2} \partial_z\Bigr)\,.\label{eq:Sframe}
\end{align}
By analogy to the ZAMO case for SOs we define, $\vS :=({{\mathbf e}\Z S^1}\cdot{\mathbf U})/({{\mathbf e}\Z S^0}\cdot{\mathbf U})$ and $\gamma\Z S:= {{\mathbf e}\Z S}^0\cdot{\mathbf U}$, so that
\begin{equation}
    \label{eq:vSrelation}
    \vS (r,z) =\gamma_\mathrm{S}^{-1}e^{-\Phi/c^2}r\Omega\,.
\end{equation}
SO and ZAMO don't measure the same velocities for the dust, indeed, we have
\begin{equation}
\label{eq:diffvZvK}
   \vD :=\frac{{{\mathbf e}\Z S^1}\cdot{{\mathbf e}\Z Z^0}}{{{\mathbf e}\Z S^0}\cdot{{\mathbf e}\Z Z^0}} =\frac{g_{\phi\phi}}{r}\chi\,,  
\end{equation}
where we call $\vD$ the dragging speed. By \eqref{eq:vZRelation}, \eqref{eq:vSrelation} and \eqref{eq:diffvZvK} we find
\begin{equation}
    \label{eq:vStovZ}
     e^{2\Phi/c^2}\gamma\Z S \vS=-H \,\gamma_\mathrm{Z}^2 \,(\vZ + \vD)\,.
\end{equation}
In particular, in the low-energy régime, the considered velocities -- i.e., $\vZ,\, \vS$ and $\vD$ -- are sub-relativistic, and thus the effective Lorentz factors $\gamma_Z\approx\gamma_S=1+\mathcal{O}(v^2/c^2)$  give higher-order corrections\footnote{We consider the low-energy régime as the limit for which all the considered speeds $v_S$, $v_Z$, $v_D$ are sub-relativistic, i.e. are less or equal to a certain reference order of magnitude $v$, tens or hundreds of kilometres per second, so that all our equations can be expanded at the lowest orders with respect to $v/c \approx 10^{-3}$.}. Given, sub-relativistic frame dragging and weak pseudo-Newtonian potential, $\Phi \approx \mathcal{O}(v^2/c^2)$, we find $A = -rv_D/c^2 + \mathcal{O}(v^2/c^2)$. Therefore, the factor $H$ can be neglected, we read from \eqref{eq:Hnorm} that it is given by $H = -1 +  \mathcal{O}(v^2/c^2)$. Thus, \eqref{eq:vStovZ} takes the approximated form
\begin{equation}
\label{eq:FirstAppVel}
    \vS = \vZ + \vD + \mathcal{O}(v^3/c^3) \, \,
\end{equation}
where $\vS \approx r\Omega + \mathcal{O}(v^2/c^2) =: \vK$ which we name as the kinetic velocity. Thus, we have
\begin{equation}
    \label{eq:VelImportant}
    \vZ = \vK - \vD + \mathcal{O}(v^3/c^3)
\end{equation}
Eq. \eqref{eq:VelImportant} clearly shows the difference between the physical velocities measured by ZAMO and SO for the dust, given by the dragging speed, even in a low-energy régime. We note that the role of ZAMO was already stressed in \cite{crosta2020, Astesiano:2021ren, Re2023, AstesianoRuggiero1,Crosta2023}, whilst the role of SO has been previously suggested in \cite{Costa1}. However, here we further elucidate their role, as this plays a crucial role in understanding the observational effects of GR within the systems under consideration.

Given the differences in dust velocities measured by different observers, we must identify the speed which is then plotted as \lq\lq galaxy rotation curve\rq\rq~for distant galaxies. We should hence wonder what physical quantity is actually measured by these curves. For any such distant galaxy \footnote{The case of the Milky Way (MW) is different because of the far smaller distances of its stars from us, that allow more refined measurement techniques. The GAIA satellite is indeed able to measure the transversal component of the stars' velocities with respect to us, appreciating their angular movements. Such a geometrical measure is profoundly different from the Doppler shift, and we will not cover it in this article.}, the speed of the stars and gas is inferred from Doppler shift $z$ measurements, usually interpreted through the special-relativistic formula
\begin{equation}
\label{SR redshift}
    1+z=:\frac{1+(v/c)\cos\theta }{\sqrt{1-v^2/c^2}} \, ,
\end{equation}
where $v$ represents the relative velocity between observer and source, and we define $\theta$ as the angle between the line of sight and the Killing vector $-\partial_{\phi}$, so that for $\theta=0$ the photon is emitted in the opposite direction of the star's motion and the photon's frequency is redshifted $z>0$. If the spatial-alone geometry is almost flat, as we expect in our case, this is approximately the angle by which the observed galaxy is flipped with respect to our line of sight, so that $\theta=0$ for a face-on galaxy, and $\theta=\pm\pi/2$ for an edge-on galaxy, with the sign changing from the left to the right side.

In the general relativistic context, the calculation of the Doppler shift is more complicated. Here we will follow the same framework introduced in \cite{Re2023} and later adopted by \cite{AstesianoRuggiero1}. Let us consider an ideal situation, with an observer at Minkowskian spatial infinity at rest with respect to the centre of the galaxy. A particle of matter inside the galaxy emits a photon with four-momentum $k_{\mu}$, which we describe in the eikonal approximation (see \cite{Misner:1974qy}), i.e. as a field
\begin{equation}
    \Psi(x^{\mu})=\Psi_0 e^{i\varphi} \quad s.t. \quad d\varphi=k_{\mu}x^{\mu}.
\end{equation}
Moreover, we will neglect here the relativistic aberration, so that the direction of the photon's motion is approximately considered to be constant along the geodesics. When the observer detects the light, it measures a Doppler shift of
\begin{equation}
    1+z:=\frac{E_e}{E_d}=\frac{1}{\sqrt{-H}}\left(1+\frac{\Omega b}{c}\right) \, .
\end{equation}
Here $E_e:=-U^{\mu}k_{\mu}$ is the emitted energy, $E_d:=-k_t$ is the detected one, and the impact parameter is $b=c k_{\phi}/k_t=-cL/E$. We notice that $-k_t=E$ and $k_{\phi}=L$ are the energy and the angular momentum of the photon, which are conserved along the geodesic given stationarity and axisymmetry of the metric.

To express these in terms of measurable quantities, it's useful to consider the spatial three-dimensional momentum $\textbf{k}$. Such a spatial projection can be performed with respect any observer, e.g., ZAMO and SO. Since we are here interested in evaluating $k_t=(\partial_t)^{\mu}k_{\mu}$, \break we perform the projection with respect to SO, i.e., ${\mathbf e}\Z S^0=e^{-\Phi/c^2}\partial_t =: u_S^\mu\partial_\mu$. We thus have to consider the projection of the photon momentum over the three-dimensional manifolds $\Sigma^3$ with Riemannian metric $h^S_{\mu\nu}:=g_{\mu\nu}+u_{\mu}^S u_{\nu}^S$, so that
\begin{equation}
\label{eq:spatialMetric}
h^S_{\mu\nu}\dd x^{\mu}\dd x^{\nu}=e^{-2\Phi/c^2}r^2 \dd\phi^2 +e^{\mu/c^2}(\dd r^2+\dd z^2) \, .
\end{equation}
Notice that, according to SO, the photon has an energy $E_S:=-u_S^{\mu}k_{\mu}=e^{-\Phi/c^2}E$. This changes along the geodesics with respect to the constant $E$ exactly because of the gravitational redshift.
Now, the four-momentum of the photon can be decomposed as $k^{\mu}=E_S(u_S^{\mu}+u_\parallel^{\mu})$, where $u_\parallel^\mu$ is a four-vector parallel to $\Sigma^3$ such that $u_\parallel^\mu\partial_\mu:=\textbf{u}_\parallel=u_\parallel^{\phi}(A\partial_t+\partial_{\phi})+u_\parallel^r\partial_r+u_\parallel^z\partial_z$. From the null condition, $k^{\mu}k_{\mu}=0$, we can check that the three-velocity of the photon has constant absolute value $|\textbf{u}_\parallel|=c$. The three-momentum is thus defined as $\textbf{k}:=c^{-2} E_S\textbf{u}_\parallel$ and has intensity $|\textbf{k}|=E_K/c$. We have defined $\theta$ as the angle between the photon's trajectory $\textbf{k}/|\textbf{k}|$ and the direction $-\hat{\phi}$, i.e. $-\cos\theta:=h_{ij}(\hat{\phi})^i k^j/|\textbf{k}|$, and thus the unit vector parallel to $\partial_{\phi}$ is
\begin{equation}
    \hat{\phi}:=\kappa\partial_{\phi} | 1=|\hat{\phi}|^2=\kappa^2 h_{\phi\phi} \Rightarrow \hat{\phi}=(h_{\phi\phi})^{-1/2}\partial_{\phi} \, ,
\end{equation}
and we find
\begin{equation}
    \cos\theta=-h_{\phi\phi}(h_{\phi\phi})^{-1/2}\frac{k^{\phi}}{|\textbf{k}|} \Rightarrow k^{\phi}=-(h_{\phi\phi})^{-1/2}|\textbf{k}|\cos\theta=-\frac{e^{\Phi/c^2}}{r}\frac{E_K}{c}\cos\theta=-\frac{E}{cr}\cos\theta.
\end{equation}
Since we need the ratio $L/E$, we explicit
\begin{equation}
    k^{\phi}=-g^{t\phi}E+g^{\phi\phi}L=\frac{g_{\phi\phi}\chi}{c^2 r^2}E +\frac{e^{2\Phi/c^2}}{r^2}L,
\end{equation}
where the contravariant components are calculated as
\begin{equation}
    g^{\phi\phi}=\frac{g_{tt}}{g_{tt}g_{\phi\phi}-g_{t\phi}^2}\,, \quad g^{t\phi}=-\frac{g_{t\phi}}{g_{tt}g_{\phi\phi}-g_{t\phi}^2}\,, \quad g_{tt}g_{\phi\phi}-g_{t\phi}^2=-c^2 r^2\,.
\end{equation}
We can thus solve for
\begin{equation}
    b=-c\frac{L}{E}=\frac{c}{g^{\phi\phi}}\left(\frac{\cos\theta}{cr}+g^{t\phi}\right)=e^{-2\Phi/c^2}\left(r\cos\theta +\frac{1}{c}g_{\phi\phi}\chi\right)
\end{equation}
obtaining for the Doppler shift
\begin{align}
\label{eq:redshift}
    1+z&=(-H)^{-1/2}\left[1+e^{-2\Phi/c^2}\frac{\Omega}{c}\left(r\cos\theta +\frac{1}{c}g_{\phi\phi}\chi\right)\right] \\
    & =\frac{1}{\sqrt{-H}}\left[\left(1+ e^{-2\Phi/c^2}\frac{v_K v_D}{c^2}\right) +e^{-2\Phi/c^2}\frac{v_K}{c}\cos\theta\right] \, . 
\end{align}
From \eqref{eq:Hnorm}, in the low-energy régime, we find 
\begin{equation}
    \label{eq:Happrox}
    -H= 1+c^{-2}(2\Phi-2\vK\vD-\vK^2) + \mathcal{O}(v^3/c^3)\,,
\end{equation}
so that \eqref{eq:redshift} becomes
\begin{equation}
\label{eq:redshift2}
    1+z = 1+\frac{v_K}{c}\cos\theta+c^{-2}\left(\frac{\vK^2}{2}-\Phi+2\vK\vD\right)+ \mathcal{O}(v^3/c^3)\, .
\end{equation}
By comparing \eqref{eq:redshift2} with \eqref{SR redshift}, we read that at the first order, the inferred velocity is the kinetic velocity, $\vK$. This provides the physical interpretation of the speed inferred through redshift measurements by SO for distant galaxies: it is essentially the same speed we plot as the galaxy rotation curve, through the Doppler shift technique.  We stress again that this is not the ZAMO speed $v_Z$, and their difference is given dragging speed $\vD$, as by \eqref{eq:diffvZvK}. 

This result allows us to better understand how the GR paradigm might bring about a reduction in the need for DM. The required amount of mass in the galaxy is usually calculated directly from the rotation curve, but such inferred speed is not proportional to the angular momentum of the matter which can be expected to be justified by the gravitational attraction of the mass itself. Rather, the angular momentum is proportional to the ZAMO speed, and can hence be smaller than what is expected from the inferred velocity $\vK$, since these exhibit a difference due to the dragging speed. We can thus expect a reduction of the required matter, and hence of the needed DM, proportionally to the amount of dragging in the galaxy.

\section{Effective dark matter contribution to galaxy rotation curve}\label{sec:Estimation}
From \eqref{eq:eHdensity}, \eqref{eq:eta}, \eqref{eq:Happrox}, and \eqref{eq:VelImportant}, the matter density on the galactic plane can be expressed at the lower order as
\begin{equation}
\label{eq:densityapp}
    8\pi G\rho(r, 0) = 2\frac{\dd\, (\vK^2)}{r \dd r}-\frac{\dd\, (r\vD)}{r \dd r}\frac{\dd\,(2r\vK-r\vD)}{r \dd r}\, + \mathcal{O}(v^4/c^4).
\end{equation}
We thus find that the required matter density to maintain the self-gravitating system in rotation is reduced in the presence of a dragging vortex rotating in the same direction of $\vK$. The reduction is not a higher-order correction if $\vD$ is a strong dragging, i.e., if it is of the same order of magnitude as $\vK$. In particular, the void condition $\rho=0$ is reached for $\eta_{,r}(r, 0)\approx \pm r^2\Omega_{,r}(r, 0)$, or equivalently for $\dd\,(r\vD)/\dd r\approx 2\vK$ and $\dd\,(r\vD)/\dd r\approx 2r\vKr$.

In \eqref{eq:densityapp} we distinguish two different, first-order, non-Newtonian effects given by the dragging: $\rho_{GR}=\rho_N+\rho_I-\rho_{II}+\mathcal{O}(v^4/c^4)$. The first goes as the square of the dragging speed and increases the required density by $+\rho_I$. This is given by
\begin{equation}
    8\pi G \rho_I = +\left[\frac{\dd \,(r\vD)}{r \dd r}\right]^2\, .
\end{equation}
The second contribution affects the measures we perform to get the rotation speed and the relation it has with the gravitational acceleration. This is proportional to the dragging speed and can reduce the density as $-\rho_{II}$ and corresponds to
\begin{equation}
    -8\pi G\rho_{II} = -2\frac{\dd \,(r\vK)}{r \dd r}\frac{\dd \, (r\vD)}{r \dd r}\, ,
\end{equation}
If the space-time dragging is strong, but not too strong (which means $\vD<2\vK$), then $\rho_{II}$ dominates and the self-gravitating system requires less mass to justify the same observed rotation curve $\vK$. 

To further elaborate on these two distinct dragging effects, we can turn to the non-linear gravitomagnetism formalism. Remarkably, from \eqref{eq:nlGM} we could deduce that the matter density required for the self-gravitating system would only be \emph{increased} as a consequence of non-linearity; instead of being decreased \cite{Costa2}. This would follow for, even if the gravitomagnetic field $\mathbf{H}$ were to be relevant at the non-linear term $H^2$, it would still appear with an opposite sign with respect to $\rho$. However, this conclusion would come only from a naive exploration of the physical system. Indeed, it all depends on what is ``kept fixed" when we ``turn on" the dragging. The non-linear gravitomagnetic equations \eqref{eq:nlGM} tell us that $\rho$ is increased by the presence of a non-negligible $\textbf{H}$ \emph{when $\textbf{G}$ has the same value}. In other words, if we were able to measure the gravitoelectric field $\textbf{G}$, our measure would require more or less $\rho$ to be explained whether $\textbf{H}$ is more or less strong. But none of our techniques can return us directly the quantity $\textbf{G}$. We measure a Doppler shift $z$, from which we inferred a velocity $v$, and then $|\textbf{G}|=v^2/r$ is calculated according to Newtonian dynamics as a centripetal acceleration. However, such a law for centripetal accelerations does not hold in GR, especially if a strong dragging is present. To make a comparison to the empirical data, we must instead "keep fixed" what is the measured quantity, i.e. the inferred speed $\vK$. Then, if we "turn on" a strong dragging $\vD$, for a fixed (inferred) profile of $\vK$, the matter density required to maintain the self-gravitation is described by \eqref{eq:densityapp}, and it is thus reduced. In particular, the second dragging effect previously discussed reduces indeed the required matter when $\dd\,(r\vD)/\dd r>0$, which means that $H_z=-H(r, z)<0$: thus why we defined $H$ with the minus sign. 

We can now move to estimating the strength of $\vD$ needed for the GR paradigm to give non-negligible DM-like contributions. To start, let us consider the Newtonian prevision of matter distribution $\rho_N=\rho_B+\rho_{DM}$, where the DM component is added to the baryonic one to explain the observed rotation curve $v$ with Newtonian dynamics. We want to reduce the DM requirement by at least some relevant fraction so that it is meaningful to speak of a true DM and some phantom part, i.e., $\rho_{DM}=\rho_{tDM}+\rho_{pDM}$. The phantom fraction should then be explained by a suitable dragging vortex, with profile $\vD(r, 0)$ that gives the desired correction, according to e.g. the formula \eqref{eq:densityapp}, where the (true) matter distribution is now $\rho =\rho_B+\rho_{tDM}$. The Newtonian model is taken in its spherical approximation, so that we assume a spherically symmetric galaxy, made up of particles in uniform circular motion. The Newtonian Equation of Motion for each of these particles hence reads
\begin{equation}
\label{eq:rhoN}
    4\pi G\, \rho_N=\frac{v^2}{r^2}+2\frac{v\,v_{,r}}{r}=\left(\frac{1}{r^2}+\frac{1}{r}\frac{\dd }{\dd r}\right)[v^2] \, ,
\end{equation}
and we can distinguish in the rotation curve the baryonic and DM contributions by $v^2=v_B^2+v_{DM}^2$. The baryonic profile can be chosen as usual to be exponentially decreasing
\begin{equation}
    \label{eq:rhoNB}
    \rho_B(r):=\rho_{B0} \, e^{-r/r_B} \, ,
\end{equation}
with parameters $r_B\cong 2.1$ kpc and $\rho_{B0}\cong 2.93\times10^{-20}\, \text{kg/m}^3$, so that the total baryonic matter is $M_B=8\pi\, r_B^3\,\rho_{B0}\cong10^{11} M_{\odot}$. In this case, the Newtonian prevision for the rotation curve becomes (if only the baryonic matter were to be present)
    \begin{equation}
    \label{eq:vNB}
    v_B(r)^2=4\pi G\,r_B^2\,\rho_{B0}\,f_B\left(\frac{r}{r_B}\right) ~| ~f_B(x)=\frac{2}{x}-\left(\frac{2}{x}+2+x\right)e^{-x}\, ,
\end{equation}
with its peak at $v_B(r_M)=v_{BM}\cong 200$ km/s, $r_M\cong 3.38363 \,r_B$. To fit an observed rotation curve that becomes flat at very large distances, that is $\lim_{r\rightarrow\infty}v(r)=v_f\approx 220$ km/s for a MW-like galaxy, we choose an isothermal DM distribution
\begin{equation}
    \label{eq:rhoNDM}
    \rho_{DM}:=\frac{\rho_{DM0}}{1+(r/r_{DM})^2}\, ,
\end{equation}
with parameters $r_{DM}\cong2.71 \, r_B\cong 5.69$ kpc and $\rho_{DM0}\cong 6.4\% \, \rho_{B0}\cong 1.88\times 10^{-21} \,\text{kg/m}^3$. The DM halo contribution to the rotation curve is then
\begin{equation}
    v_{DM}(r)^2=4\pi G\, r_{DM}^2\,\rho_{DM0}f_{DM}\left(\frac{r}{r_{DM}}\right) ~| ~f_{DM}(x)=1-\frac{\arctan x}{x} \, .
\end{equation}
This Newtonian framework is illustrated in Figure 1. The rotation curves are plotted until $50$ kpc from the centre of the galaxy.
\begin{figure}
            \label{RotCurveN}
		\centering
		\includegraphics[width = 12 cm]{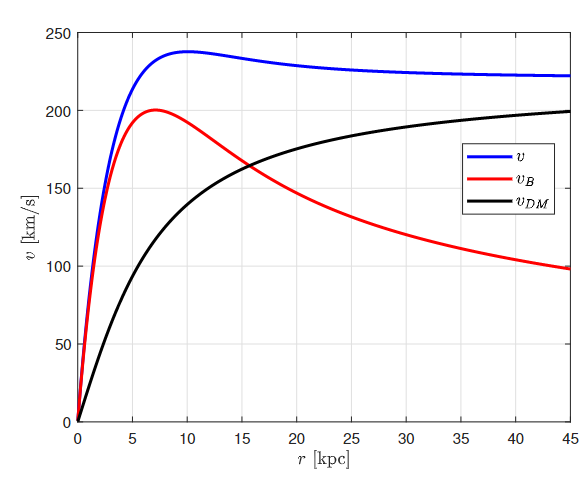}
	   \caption{The typical rotation curve $v$ of a disc galaxy. According to the Newtonian paradigm, it results from the disc contribution $v_B$ and the halo contribution $v_{DM}$.}
\end{figure}
\\
\\
\noindent 
Now, let's push forward the GR paradigm. We want to explain at least some fraction $\rho_{DM}$ as phantom DM, due to a suitable dragging profile $\vD(r,0)$. To deduce the required $\vD$, we need a GR formula that links the quantities $\vK,\, \vD$ and $\rho$. At the moment, we can exploit \eqref{eq:densityapp}, although it is valid only for the pressureless class $(\eta, H)$. Moreover, we would want to find again the Newtonian description $\rho_{pDM} = 0$ for the degenerate case $\vD = 0$. Unfortunately, for a zero dragging the \eqref{eq:densityapp} does not reduce to the Newtonian formula \eqref{eq:rhoN} that we used, for the latter assumes a spherical symmetry, whilst the Newtonian limit of the $(\eta, H)$ solutions exhibit cylindrical symmetry. This obstacle can be rigorously sidestepped only with a more refined modelling, taking into account non-zero pressure in both the Newtonian and GR galaxy, eventually in their thin-razor limit. However, for a first order estimation as the one we intend to perform in this Section, we will just add the \emph{ad hoc} strictly positive term $2\,\vK^2/r^2$ to \eqref{eq:densityapp}, so that it is forced to match \eqref{eq:rhoN} for null dragging. We thus find
\begin{equation}
\label{eq:vDphantomDM}
    \vD \approx \vK-\frac{\eta}{r} \quad s.t. \quad \frac{\dd\eta}{\dd r}\approx\sqrt{\left(\frac{\dd \, (r\vK)}{\dd r}\right)^2-8\pi G\,r^2\,\rho_{pDM}} \, .
\end{equation}
\eqref{eq:vDphantomDM} can be applied only to explain a not-too-large DM fraction $8\pi G\,\rho_{pDM}(r)\leq \left[(\dd\,(r\vK)/\dd r)/r \right]^2$. We stress that this is not a real physical limitation of a dragging solution with pressure, but only of our \emph{ad hoc} forcing.
\begin{figure}
        \label{RotCurveGR}
	\centering
	\includegraphics[width = 12 cm]{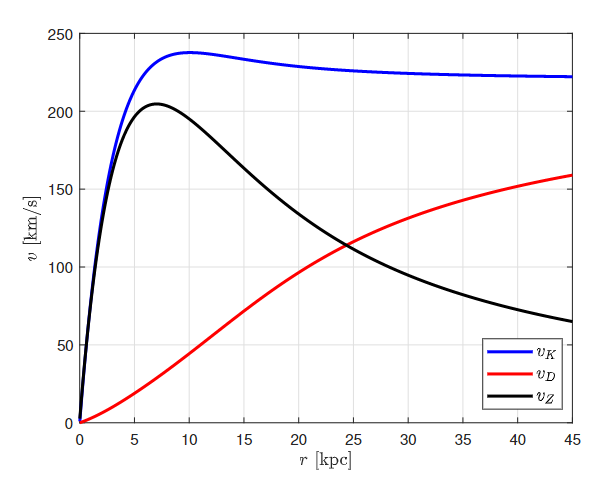}
	\caption{The same rotation curve $v_K$ of Figure 1, for which a GR study is performed. A total $50$\% of the DM inferred for the galaxy is here explained as fictitious effect, due to the strong dragging with profile $\vD$.}
\end{figure}

Under these \emph{caveat}, we apply \eqref{eq:vDphantomDM} to explain a phantom DM profile that is chosen to be proportional to the total DM one, $\rho_{pDM}=\alpha\,\rho_{DM}$, for some $0\leq\alpha\leq 1$. \eqref{eq:vDphantomDM} then becomes
\begin{equation}
    \left(\frac{\dd \eta}{\dd r}\right)^2\approx\left(\frac{\dd \,(r\vK)}{\dd r}\right)^2-2\alpha\left(v_{DM}^2+r\frac{\dd\,(v_{DM}^2)}{\dd r}\right) \, ,
\end{equation}
so that it always returns real numbers for $\alpha\leq0.5$. A simple numerical integration gives us the plots in Figure 2 for the choice $\alpha:=0.5$. We see that an asymptotic, subrelativistc value for $\vD$ of $\approx 35$ km/s is then sufficient to explain half of the DM content inferred from the galactic plane rotation curves. In particular, for this case, the value of $\vD$ for our galactic neighbourhood, i.e. for $r=r_{\odot}\cong8.6$ kpc would then depend, on $\alpha$ as
\begin{equation}
    \vDSol = \vD(r_{\odot},0)\approx \alpha\times 71\, \text{km/s} \, .
\end{equation}
We can summarize these results by saying that the rotation curves of disc galaxies can be explained, at least for some non-negligible fraction $\alpha$, by the presence of a dragging vortex with a speed $\vD\propto\alpha$ of tens of kilometers per second. It does not exceed the order of magnitude of $\vK$, as we assumed throughout this article, so that the galaxy would indeed be in a low-energy régime, as expected.

As a final remark, we note that qualitatively similar results for the impact of GR on rotation curves can be found in \cite{AstesianoRuggiero2} within the linear gravitomagnetic approach discussed at the end of Section \ref{sec:intro}. Here, we have moved a step further and obtained the results corresponding to what Astesiano and Ruggiero defined as \lq\lq strong gravitomagnetism \rq\rq in Section 4 of \cite{AstesianoRuggiero2}. Furthermore, we presented a thorough analysis of the dragging vortex when the non-linearities of GR have been properly considered.

\section{Effective dark matter contribution to gravitational lensing}\label{sec:GravLensing}

The role of frame-dragging on gravitational lensing was previously studied in literature, see e.g. \cite{Asada_2000,Sereno1, Sereno2}. However, these calculations were always performed for the linearised Einstein Equations, i.e., within the conventional gravitomagnetism approach. Here we are interested in the full GR calculation, first discussed in \cite{Costa2}, since the model developed so far shows crucial nonlinear features.

Let us consider a disc galaxy whose average dynamics is described by a metric of the type \eqref{eq:metric} with reflection symmetry with respect to the equatorial plane. Then, the velocities of the self-gravitating galactic matter on opposite sides of the galaxy are antiparallel. Therefore, in both cases the matter is pushed inwards, or outwards, by the gravitomagnetic force on both sides. However, this symmetry is broken for light deflection in gravitational lensing phenomena. Indeed, let us consider a photon in motion on the equatorial plane with four-momentum $k^{\mu}$, so that we can take its three-dimensional spatial part $\textbf{k}=E_K\,\textbf{u}$ in the same way we did in Section \ref{sec:Observers}, so that $\textbf{u},\, \textbf{H}, \, \textbf{B}$ are all spatial vectors lying on the same three-manifold $\Sigma^3$. Its three-dimensional part $\mathbf{u}$ is always directed towards us, and the field $\mathbf{H}$ is always parallel to the $z$-axis. Thus, the photon's trajectory is always pushed in the same $\mathbf{u}\times\mathbf{H}$ direction by the dragging vortex (see Figure 3). In \cite{Costa2}, from this crucial observation and qualitative calculations for the gravitational lensing on the equatorial plane, it is then argued that the presence of a dragging vortex would bring about a simple shift of the lensed images and not a further focusing of the light. Here, we prove this conclusion to be fundamentally wrong within our models. Furthermore, we note that such a conclusion is already at odds with the known effect on rotating lenses in linearised GR \cite{Sereno1, Sereno2}.
\\
\\
Consider a conjunction as the one illustrated in Figure 3. Two photons are emitted by a light source with a divergence angle $\theta_S$, travel near a galaxy whose geometry is described by an effective metric \eqref{eq:metric}, with trajectories $C_{\pm}$ bent by gravity, and arrive to us from two directions which differ for an angle $\theta_R$. We should interpret this scheme by saying that only two apparent images are generated by such gravitational lensing (instead of an Einstein ring, or an arc). If the GR paradigm is correct, the apparent angular distance between these images should be non-negligibly increased by the presence of a strong dragging, resulting in a contribution of phantom DM. The calculation of $\theta_R$ is quite difficult in almost all cases. In fact, the method relying on the use of the Gauss--Bonnet theorem used in \cite{Galoppo1} and \cite{Costa2} can be applied only if the photons' trajectories lie on a well-defined two-surface. We notice that for a photon moving in the geometry defined by \eqref{eq:metric} this is generally not the case. For example, if we consider planar trajectories, these exist only if both the light source and the observer are exactly on the galactic plane $z=0$ (which implies that the galaxy must be perfectly edge-on from our viewpoint). However, it is instructive to solve the problem in this highly idealised case. For this purpose, we can apply the Gauss--Bonnet theorem to the geometry of Figure 3 as
\begin{equation}
\label{eq:GBT}
    \theta_R+\theta_S = \int\int_{\mathcal{S}}K \, da+\int_{\partial\mathcal{S}}\kappa_g \,d\lambda +2\,\pi\,(1-\chi(\mathcal{S}))\, ,
\end{equation}
where $\mathcal{S}$ is the surface bounded by $C_{\pm}$, $\chi(\mathcal{S})=1$ is its Euler characteristic, $K$ is its Gaussian curvature, and $\kappa_g=(\mathbf{G}+\mathbf{u}\times\mathbf{H})\cdot\hat{e}_{\perp}$ is the geodesic curvature of $C_{\pm}$, calling $\hat{e}_{||},\, \hat{e}_{\perp}$ the unit vectors respectively parallel and orthogonal to the trajectories $C_{\pm}$ at each of their points. Since $|\textbf{u}|=c$, as we showed in Section \ref{sec:Observers}, and $\textbf{u}$ is defined as parallel to $\hat{e}_{||}$, we will substitute $\textbf{u}=c\hat{e}_{||}$. To properly carry out the calculation, it is  crucial to evaluate the fields $\mathbf{G}(r), \mathbf{H}(r)$ and also the quantity $\mathbf{u}(r)$ for slightly different values of $r_{\pm}(\lambda)$, which are greater for the left path $C_+$ with respect to the right path $C_-$.
\\
\begin{figure}
\label{FigureLensing}
		\centering
		\includegraphics[width=12 cm]{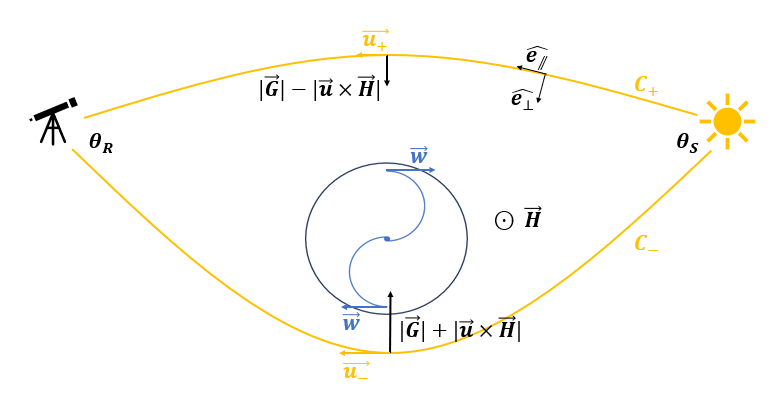}
		\caption{A sketch of gravitational lensing due to a dragging metric. The asymmetry between the co-rotating and counter-rotating side is stressed, as it is the reason why such lensing results to be bigger than the zero dragging case. Here, the galaxy is shown viewed from below (i.e., negative $z$)}
\end{figure}

\noindent
In Figure 3, we stressed that the left-right symmetry is broken in the presence of a strong dragging. We see that $C_+$ should be bent by a weaker acceleration $|G_{\perp}-(\mathbf{u}\times\mathbf{H})_{\perp}|$, while $C_-$ feels a greater centripetal acceleration $|G_{\perp}-(\mathbf{u}\times\mathbf{H})_{\perp}|$. Notice also that the resulting light paths $C_{\pm}$ are not invertible. Let us then write as a first approximation $r_{\pm}(\lambda)\approx r_0(\lambda)\pm\Delta r(\lambda)$, where $r_0(\lambda)$ describes the trajectories the photons would follow if $\mathbf{H}$ were to be null. Thus, \eqref{eq:GBT} takes the form
\begin{align}
\label{eq:GBTspecialized}
    \theta_R+\theta_S &=  \int\int_{\mathcal{S}}K da +\sum_{\pm}\pm\int_{C_{\pm}}\left[\mathbf{G}(r_{\pm}(\lambda))+\mathbf{u}(r_{\pm}(\lambda))\times\mathbf{H}(r_{\pm}(\lambda))\right]\cdot\mathbf{e}_{\perp}(r_{\pm}(\lambda)) \, d\lambda= \cr
     & = \int\int_{\mathcal{S}}K da + \int[G_{\perp}^+ -|\mathbf{u}^+\times\mathbf{H}^+|-G_{\perp}^- +|\mathbf{u}^-\times\mathbf{H}^-|]\, d\lambda \,.
\end{align}
We can also assume that the light is passing in the galaxy's periphery, where all the fields are rapidly decreasing under the assumption of Minkowskian asymptoticity at $r\rightarrow\infty$. We can thus evaluate at the first order $G(r_{\pm})\approx G(r_0)\pm G_{,r}(r_0) \, \Delta r$ and  $H(r_{\pm})\approx H(r_0)\pm H_{,r}(r_0) \, \Delta r$, where we defined the fields so that $G, \, H$ are positive, thus the derivatives $G_{,r}<0,\, H_{,r}<0$ are negative. Substituting our results in \eqref{eq:GBTspecialized} and neglecting the higher order terms, we find
\begin{align}
    \label{eq:GBTfinal}
    \theta_R+\theta_S\approx\int_{\mathcal{S}}K da+ 
    2\int\left[G(r_0)\cos\left[\alpha(r_0)\right]-cH_{,r}(r_0)\Delta r (\lambda) \right]d\lambda \, ,
\end{align}
where $\alpha(r_0)$ is the angle between $\mathbf{G}$ and $\mathbf{u}$ at $r_0$. In \eqref{eq:GBTfinal} the corrections due to $H$ are positive, thus acting as a phantom dark matter. The spatial Gaussian curvature $K$ depends only on the spatial metric \eqref{eq:spatialMetric}, where no dragging terms appear at the linear order. Moreover, the $G$ term is not affected by the presence of dragging. We can thus evaluate the first order dragging contribution as
\begin{equation}
    \Delta\theta_{dragging}\approx -2c\int H_{,r}(r_0) \Delta r(\lambda) d\lambda \approx \frac{2}{c}\int\left(\frac{\vD}{r^2}-\frac{\vDr}{r}-\vDrr\right)\Delta r (\lambda)\, \dd \lambda \, >0 \, .
\end{equation}
We can expect this contribution to be non-negligible when $\vD$ has the same order of magnitude of $\vK$ ("strong dragging"). Therefore, if the spacetime geometry for disc galaxies can be reasonably approximated by a  metric of the type \eqref{eq:metric}, we find that a strong dragging can produce DM-like effect even for gravitational lensing observables.
\section{Conclusions}\label{sec:conclusions}

The class of axysimmetric, stationary spacetimes sourced by dust -- here dubbed $(\eta,H)$ class -- represents a viable family of fully general relativistic models to describe the average disc galaxy dynamics around the galactic plane. In these models, the space-time geometry around disc galaxies is characterized by the presence of a non-negligible dragging field -- essentially the off-diagonal component of the resulting metric. This dragging vortex then generates dark matter-like effects which can starkly impact the galaxy rotation curves and gravitational lensing observables. Therefore, such fully general relativistic disc galaxy models could demand a complete revision of the amount of inferred dark matter in such systems.

In this paper, we have discussed the various formalism, and the relevant observers, for the modelling of disc galaxies in General Relativity within the $(\eta,H)$ framework. In particular, we have identified the correct observables for the study of the redshift-inferred rotation curves of distant galaxies. 

We carried out the first study of the low-energy régime of the fully $(\eta,H)$ solutions, whilst accounting for GR non-linearities, which should better be suited to describe systems such as galaxies. We have thus shown that a sub-relativistic dragging speed is sufficient to produce a non-negligible dark matter-like effect in disc galaxies. In particular, we estimated that an asymptotic dragging velocity below 50 km/s would be sufficient to explain away approximately 50\% of the dark matter inferred from the rotation curves for a Milky Way-like galaxy.

Furthermore, we have calculated the effects of non-negligible dragging on strong gravitational lensing by disc galaxies. We find that these would be severely impacted by the presence of such a dragging field. Indeed, the geometries studied produce significant dark matter-like effects that augment the bending angle of light in gravitational lensing.  

Nevertheless, we are fully aware that the controversy of whether real disc galaxies are surrounded by such vortices of dragging, and whether they account for some phantom dark matter, is far from being settled. What we have shown is that such dragging solutions are allowed by the fundamental theory of gravity as we know it. Therefore, in our opinion, the presence of a geometry-defining dragging field in disc galaxies is a viable hypothesis to explain a significant fraction of the inferred dark matter in disc galaxies, thus carrying far-reaching consequences for astrophysics and cosmology.

Finally, since the $(\eta, H)$ solutions do not incorporate any effective pressure in the modelling of the physical disc galaxy, they can not be considered global models for such systems. Therefore, our future work will focus on the introduction of pressure in galaxy modelling in General Relativity to sidestep the drawbacks of the $(\eta,H)$ class.

\emph{Acknowledgments}.
We are grateful to Davide Astesiano, Sergio Cacciatori, Massimo Dotti, Leonardo Giani, Vittorio Gorini, Christopher Harvey-Hawes, Morag Hills, Emma Johnson, Zachary Lane, Oliver Piattella, Shreyas Tiruvaskar, Michael Williams and David Wiltshire for stimulating discussions. 

\bibliographystyle{ieeetr}
\bibliography{refs.bib}

\end{document}